\documentstyle[11pt,aasms4]{article}

\newcommand\cL{{\cal L}}

\newcommand{\kms}{km \hskip -2pt s$^{-1}$}
\newcommand{\mkms}{{\rm km}\,{\rm s}^{-1}}
\newcommand{\hmpc}{\;h^{-1}\;{\rm Mpc}}
\newcommand{\dbar}{{\overline d}}
\newcommand{\rmax}{R_{\rm max}}
\newcommand{\nres}{N_{\rm res}}
\newcommand{\model}{{\rm Model}}
\newcommand{\data}{{\rm Data}}

\newbox\grsign \setbox\grsign=\hbox{$>$} \newdimen\grdimen \grdimen=\ht\grsign
\newbox\simlessbox \newbox\simgreatbox
\setbox\simgreatbox=\hbox{\raise.5ex\hbox{$>$}\llap
     {\lower.5ex\hbox{$\sim$}}}\ht1=\grdimen\dp1=0pt
\setbox\simlessbox=\hbox{\raise.5ex\hbox{$<$}\llap
     {\lower.5ex\hbox{$\sim$}}}\ht2=\grdimen\dp2=0pt

\begin{document}
\title{THE TOPOLOGY OF LARGE SCALE STRUCTURE 
IN THE 1.2 JY IRAS REDSHIFT SURVEY}

\author{Zacharias A.M. Protogeros}
\affil{Ohio State University, Department of Physics, Columbus, OH 43210}

\author{David H. Weinberg}
\affil{Ohio State University, Department of Astronomy, Columbus, OH 43210}

\affil{E-mail: zack@ohstpy.mps.ohio-state.edu, dhw@payne.mps.ohio-state.edu}

\begin{abstract}
We measure the topology (genus) of isodensity contour surfaces in 
volume limited subsets of the 1.2 Jy IRAS redshift survey,
for smoothing scales $\lambda=4\hmpc$, $7\hmpc$, and $12\hmpc$.
At $12\hmpc$, the observed genus curve has a symmetric form
similar to that predicted for a Gaussian random field.
At the shorter smoothing lengths, the observed genus curve shows
a modest shift in the direction of an isolated cluster or
``meatball'' topology.  We use mock catalogs
drawn from cosmological N-body simulations to investigate the
systematic biases that affect topology measurements in samples
of this size and to determine the full covariance matrix of the
expected random errors.
We incorporate the error correlations into our evaluations of
theoretical models, obtaining both frequentist assessments of
absolute goodness-of-fit and Bayesian assessments of
models' relative likelihoods.
We compare the observed topology of the 1.2 Jy survey to the
predictions of dynamically evolved, unbiased, gravitational instability
models that have Gaussian initial conditions.
The model with an $n=-1$, power-law initial power spectrum achieves the 
best overall agreement with the data, though models
with a low-density cold dark matter power spectrum
and an $n=0$ power-law spectrum are also consistent.
The observed topology is inconsistent with an initially Gaussian
model that has $n=-2$, and it is strongly inconsistent with
a Voronoi foam model, which has a non-Gaussian, bubble topology.
\end{abstract}


\section{Introduction}

According to the most popular theories of structure formation, the observed
distribution of galaxies --- a complex network of clusters, superclusters,
tunnels, and voids --- developed by gravitational instability from Gaussian
primordial fluctuations.  Two different and complementary approaches have
been followed to test the Gaussian hypothesis.  The first uses the
probability distribution function (PDF) or its moments 
(skewness, kurtosis, etc.);
observed results are compared to predictions for 
a gravitationally evolved Gaussian field, which are computed
either by numerical simulations or by various approximation schemes
(e.g., Fry 1984; Bernardeau 1992; Juszkiewicz, Bouchet, \& Colombi 1993;
Juszkiewicz et al. 1995; Bernardeau \& Kofman 1995; 
Protogeros \& Scherrer 1997).
The second approach uses topological characteristics of the galaxy density
field, quantified by percolation analysis (Shandarin \& Zel'dovich 1983;
Yess, Shandarin \& Fisher 1997) or by the genus of
isodensity contours (Gott, Melott, \& Dickinson 1986, hereafter GMD;
Gott, Weinberg, \& Melott 1987, hereafter GWM; for a review see
Melott 1990).  The genus measure also yields constraints on the index
of the primordial power spectrum by quantifying the ``corrugation'' of
structure in the smoothed density field.  In this paper we apply
the genus method to one of the largest complete galaxy redshift
surveys, the 1.2 Jy IRAS redshift survey (Fisher et al. 1995).
We make extensive use of mock catalogs drawn from cosmological N-body
simulations to estimate systematic and random errors and to evaluate
the viability of models.  Our statistical methodology should
also be useful for topological analyses of future, larger redshift surveys.

Our basic approach follows that of GMD, GWM, and Gott et al. (1989).
From the galaxy distribution, we create a density field by convolving
with a Gaussian window, 
\begin{equation}
W(r)={1\over {(2 \pi)^{3/2}\lambda^3}}e^{-r^2/{2\lambda^2}}.
\label{window}
\end{equation}
[Note that our definition of the smoothing length, $\lambda$, 
based on the conventional
form of a Gaussian window, differs by a factor of $2^{1/2}$ from that
used in the Gott et al. papers.]
We then construct isodensity contours at a variety of threshold levels
and measure the genus of each.  Applying the Gauss-Bonnet theorem,
GMD define the genus $G_s$ through the integrated Gaussian curvature,
\begin{equation}
G_s\equiv -{1\over {4\pi}} \int_S K dA,
\label{curvature}
\end{equation}
with
$K \equiv 1/(a_1 a_2)$ where $a_1$ and $a_2$ are the principal
radii of curvature.
Equation~(\ref{curvature}) differs slightly from the standard mathematical
definition of the genus, 
but it is useful for cosmological purposes because it can
be applied to a contour that runs into the boundary of a finite survey
and it defines a quantity that is, statistically, proportional to volume.
For a compact surface, $1+G_s$ is the number of handles or holes
(in the sense of donut holes), while a surface broken into $n$ disjoint,
simply connected pieces (e.g., $n$ spheres) has $G_s=-n$.
We measure $G_s$ using the program CONTOUR (Weinberg 1988), based on
the algorithm of GMD (see Coles, Davies, \& Pearson 1996 for an
alternative method of computing the genus).

For a Gaussian random field, the mean genus per unit volume is
\begin{equation}
g_s=A(1-\nu^2)e^{-\nu^2/ 2},
\label{wcurve}
\end{equation} 
where $\nu$, is the threshold density of the contour
in units of the standard deviation (Doroshkevich 1970; Adler 1981;
Bardeen et al. 1986; Hamilton, Gott, \& Weinberg 1986).
Positive and negative fluctuations
are statistically interchangeable in a Gaussian field, so the
$\nu=0$ contour has a spongelike topology (positive $G_s$) in which
the high and low density regions are both multiply connected and 
mutually interlocking.  At high or low $\nu$ the genus becomes negative
as a typical contour breaks into separate bags around isolated clusters
or voids, but the dependence is symmetric about $\nu=0$.
The normalizing constant $A$ depends on the second
moment of the power spectrum.  For a field with a power-law 
spectrum $P(k) \propto k^n$ smoothed with the Gaussian filter of
equation~(\ref{window}), it is
\begin{equation}
A= {1 \over 4\pi^2\lambda^3} \left( { {3+n}\over{6}} \right)^{3/2}.
\label{amplitude}
\end{equation}
A field with more small scale power (higher $n$) has choppier, more
corrugated structure, and hence a higher genus per unit volume for a given
smoothing length.  Since the smoothing length provides the only characteristic
length scale in a Gaussian field with a power-law spectrum, the mean
genus per unit volume necessarily scales as $\lambda^{-3}$ for fixed $n$.

Linear evolution preserves the Gaussian character of the initial density
field.  Nonlinear evolution does not, but the effects of 
nonlinear evolution on the genus curve are modest if the smoothing length
is greater than or equal to the correlation length {\it and} one 
characterizes contours by the fractional volume they enclose rather
than the density level {\it per se} (GWM; Weinberg, Gott \& Melott 1987).
Volume weighting also makes the genus curve insensitive to ``biased''
galaxy formation, since even nonlinear bias tends to maintain a monotonic
relation between galaxy density and mass density.
The information ``lost'' by volume weighting is precisely that contained
in the PDF, so with this approach the genus curve and PDF at a given
smoothing scale provide independent and complementary information about the
density field.  For convenience, we characterize
a contour that encloses fractional volume $f$ (in the region above
the threshold density) by the value of $\nu$ for a corresponding
contour in a Gaussian field, defined through the implicit equation
\begin{equation}
f = (2\pi)^{-1/2} \int_\nu^\infty e^{-x^2/2} dx.
\label{nudef}
\end{equation}
With this definition, equation~(\ref{wcurve}) continues to hold for a
Gaussian field, and it remains a good first approximation as the field
evolves into the nonlinear regime (Melott, Weinberg, \& Gott 1988;
Park \& Gott 1991).
Using second-order perturbation theory, Matsubara (1994) and
Matsubara \& Suto (1996) show that 
even weakly non-linear evolution distorts the shape of the genus curve 
if it is plotted as a function of density contrast rather than 
fractional volume or equivalent $\nu$.

The genus statistic has been applied previously to six different
redshift surveys of optically selected galaxies
(Gott et al. 1989; Park, Gott, \& da Costa 1992;
Vogeley et al. 1994), and to the QDOT survey,
a 1-in-6 subset of IRAS galaxies with $60\mu$ flux density
brighter than 0.6 Jy (Moore et al. 1992).
It has also been applied to redshift surveys of Abell clusters
(Gott et al. 1989; Rhoads, Gott, \& Postman 1994), and its 2-dimensional
analog has been applied to redshift slices (Park et al. 1992;
Colley 1996),
to projected galaxy and cluster catalogs (Coles \& Plionis 1991;
Plionis, Valdardini, \& Coles 1992; Gott et al. 1992),
and to COBE maps of cosmic microwave background fluctuations
(Smoot et al. 1994; Colley, Gott, \& Park 1996; Kogut et al. 1996).

The 1.2 Jy IRAS survey contains 5321 galaxies and covers all of the sky except
for the Galactic plane (Galactic latitude $|b|<5^\circ$) and a number
of small, isolated patches at high Galactic latitude, covering about
4\% of the sky in total.  For our analysis we use a catalog provided
by M.\ Strauss in which these high latitude regions
have been filled with randomly placed galaxies.  Observational details
of the survey are described in Strauss et al. (1992) and Fisher et al. (1995).
The survey has been the basis for many statistical investigations
of large scale structure including the power spectrum
(Fisher et al. 1993; Cole, Fisher, \& Weinberg 1995), the
two-point correlation function (Fisher et al. 1994ab), and moments of
the counts-in-cells distribution (Bouchet et al. 1993).
It has also been used in comparisons between predicted and observed
peculiar velocity fields (e.g., Davis, Nusser, \& Willick 1996).

For topological analysis we use volume limited subsets of the redshift
survey, so that the physical properties of the tracer galaxies and the
effects of shot noise are uniform throughout the survey volume.
A volume limited sample consists of those galaxies within distance
$\rmax$ that are luminous enough that they would still exceed the survey
flux limit if they were at distance $\rmax$.  The number of galaxies
in a volume limited sample first increases with $\rmax$ as the survey volume 
grows, then declines at large $\rmax$ as the fraction of galaxies luminous
enough to be seen at the sample edge begins to decline rapidly.
For the 1.2 Jy survey, the size of a volume limited sample peaks at
$\sim 1100$ galaxies for $\rmax \approx 60\hmpc$.$^1$
\footnotetext[1]
{$h \equiv H_0 / 100 \,\mkms\,{\rm Mpc}^{-1}$.}
We infer a galaxy's distance from its redshift referred
to the frame of the Local Group.

The smoothing length must be large enough to suppress shot noise fluctuations
in the density field, but at fixed $\rmax$ increasing the smoothing
length reduces the number of independent resolution elements
in the survey volume.  Following the rule of thumb suggested
by Weinberg et al. (1987) and used in subsequent observational analyses,
we adopt a smoothing length $\lambda \approx \dbar/\sqrt{2}$,
where $\dbar \equiv n_g^{-1/3}$ is the mean intergalaxy separation.
(The $\sqrt{2}$ factor does not appear in the earlier papers because
of their different Gaussian filter definition.)
We discuss this choice further in \S 3 below.  
The number of resolution elements in the smoothed density field,
i.e., the ratio of the survey volume to the smoothing volume, is
\begin{equation}
\nres= {\omega_s \rmax^3 \over 3}\, {1 \over (2 \pi)^{3/2} \lambda^3},
\label{nres}
\end{equation}
where $\omega_s = 4\pi(1-\sin{\pi \over 36})$ sr is the solid angle
of the 1.2 Jy survey.

\begin{figure}
\centerline{
\epsfxsize=3.5truein
\epsfbox[105 445 515 680]{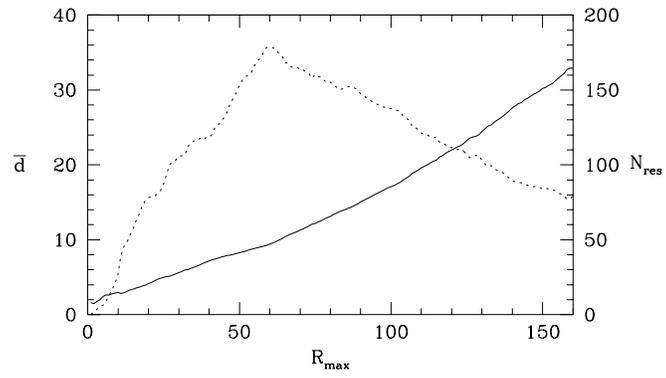}
}
\caption{\protect\small The mean separation $\bar d$
(solid line, left axis scale) and the number of resolution elements 
$\nres$ (dotted line, right axis scale) for volume limited samples of 
depth $\rmax$.  Distances are in $\hmpc$.
\vskip 2.0truein
\null
\label{fig:nres}
\normalsize}
\end{figure}

Figure~1 plots the mean separation $\dbar$ and the number of resolution
elements $\nres$, computed from equation~[\ref{nres}] assuming
$\lambda=\dbar/\sqrt{2}$, as a function of the sample depth $\rmax$.
The number of resolution elements (which is proportional to the
number of galaxies in the volume limited sample) peaks at $\sim 170$ for
$\rmax=60\hmpc$, where the mean separation is $\dbar=9.5\hmpc$.
In this sense, the $\rmax=60\hmpc$ sample is the optimal volume limited
subset that we can construct for topological analysis, and we focus
our greatest attention on this sample.  The corresponding smoothing
scale is $\lambda \approx 7\hmpc$.
Since the dependence of topology on smoothing scale is itself
interesting, we also analyze samples with $\rmax=30\hmpc$ and
$100\hmpc$, using smoothing scales $\lambda = 4\hmpc$ and $12\hmpc$,
respectively.

In the next section we describe our procedure for creating mock catalogs 
designed to mimic the 1.2 Jy redshift survey.  In \S 3 we use these
mock catalogs to study the systematic distortions in the genus curve that
arise from shot noise, the finite size of the survey volume, and
peculiar velocities.  In \S 4 we use the mock catalogs to examine the
magnitude and covariance of random errors in the genus curves of the
1.2 Jy subsamples, and we 
outline a statistical methodology for comparing observed and predicted
genus curves.  In \S 5 we present results for the 1.2 Jy survey and 
compare them to theoretical predictions.  We summarize our conclusions
in \S 6.

\section{Construction of Mock Catalogs}
In order to assess uncertainties in our observational analysis and
generate theoretical predictions for comparison to the data, we
want realistic mock catalogs that will be subject to the same
systematic effects as the 1.2 Jy survey.  For our primary set of mock
catalogs, we use the same N-body simulations as Cole et al. (1995).
These simulations have Gaussian initial conditions with the 
$\Gamma=0.25$ power spectrum of Efstathiou, Bond \& White (1992),
which produces large scale clustering consistent with recent studies
of IRAS galaxies.  The linear theory power spectrum is normalized
to an rms fluctuation $\sigma_8=0.8$ in spheres of radius $8\hmpc$
at $z=0$, and the density parameter is $\Omega_0=0.3$
(with no cosmological constant).
There are four independent realizations, each of 
a periodic box of size $l_{\rm box} =400 \,\hmpc$. 
The simulations use a staggered-mesh PM code written by C. Park (1990)
to evolve a density field represented by $200^3$ particles, 
with a $400^3$ mesh for force computations.
The simulations begin at a redshift of 24 and evolve
to the present in 48 equal steps, using the expansion factor
$a$ as the time variable for integration.
The large timesteps are adequate because
of the rather low ($\sim 1-2 \hmpc$) force resolution of the computations. 
Because we are interested in properties of the galaxy density 
field smoothed over several $\hmpc$, this force resolution is more
than sufficient for our purposes.

To create a mock catalog from the simulation, we first select a random
particle to represent the ``observer.''  The Local Group is known to
be in a region where shear and dispersion in the peculiar velocity field
are fairly low.  We therefore reject observer particles if the 
$3-$dimensional peculiar velocity dispersion within
a surrounding sphere of radius $5\hmpc$ exceeds 200 \kms, thus avoiding
observers in rich clusters and other regions where peculiar velocity
distortions would be radically different from those affecting samples
centered on the Local Group.
Given an observer that passes this velocity dispersion cut, we select
particles in a surrounding sphere of radius $\rmax$.
We compute redshift space positions 
${\bf r}=H_0({\bf v}-{\bf v_0})\cdot {\bf \hat{r}}$,
where ${\bf v}$ is the particle velocity (including Hubble flow relative
to the observer) and ${\bf v_0}$ is the average velocity of all 
particles within $1\hmpc$ of the observer, thus mimicking the procedure
of referring galaxy redshifts to the Local Group frame.
We randomly sample the particle distribution within the sphere
to obtain a mean interparticle separation equal to that of the
corresponding IRAS sample: $\dbar=9.5\hmpc$ for $\rmax=60\hmpc$,
$\dbar=5.5\hmpc$ for $\rmax=30\hmpc$, and $\dbar=17.0\hmpc$ for
$\rmax=100\hmpc$.  Finally, we eliminate particles in a 10-degree wedge
to represent the Galactic plane cut in the 1.2 Jy survey.
We create 512 mock catalogs for each value of $\rmax$,
drawing 128 from each of the four simulations.  The ratio of the
volume of an individual simulation to the volume of an individual
mock catalog is $77 \times (60\hmpc / \rmax)^3$, so in general these
128 mock catalogs are largely but not completely independent.

The $\Gamma=0.25$ power spectrum has the shape predicted for a 
cold dark matter (CDM) model with scale-invariant primeval fluctuations,
$\Omega_0=0.3$, and a Hubble constant $h \sim 0.8$ (Efstathiou et al. 1992).
We therefore refer to these simulations below as the CDM simulations.
In order to test Gaussian models with other initial power spectra,
we also construct mock catalogs from simulations with initial
spectra $P(k) \propto k^n$ with $n=0$, $-1$, and $-2$.  We again assume
$\Omega_0=0.3$, normalize the linear power spectra to $\sigma_8=0.8$,
and run four realizations of each model.  The numerical
parameters of these simulations are similar to those of the CDM 
simulations, except that they use $150^3$ particles, a $300^3$ mesh,
and a cube of size $300\hmpc$.  In all of our models we choose a random
subset of N-body particles to represent galaxies, thus implicitly assuming
that galaxies are unbiased tracers of the mass.

\section{Systematic Effects}
When measuring the genus curve of a mock catalog or of a volume
limited subset of the 1.2 Jy survey, we first compute the galaxy
density field on a cubic mesh, using cloud-in-cell (CIC) binning.
We convert this density field $\rho_g$ to a density contrast field
$\delta_g = (n_g-{\overline n}_g)/{\overline n}_g$,
with ${\overline n}_g = N_g/V_s$ the mean galaxy density of the
sample.  We set $\delta_g=0$ in all mesh cells outside the
sample volume, i.e., with $R>\rmax$ or within $5^\circ$ of the
Galactic plane.  We also create a ``mask'' array that is 1.0 for all
cells within the sample volume and 0.0 for all exterior cells.
We smooth the galaxy density contrast field by convolving it with
the Gaussian window function~(\ref{window}).  At each cell, we want
the convolution to cover only that portion of the smoothing window
that lies within the sample volume.  Technically, we accomplish this
objective by smoothing both the density contrast array and
the mask array with a Fast Fourier Transform convolution, then
dividing the smoothed density contrast by the smoothed mask.
Since $\delta_g=0$ outside the sample boundary, the exterior regions do not
contribute to the convolved density contrast, and dividing by the
smoothed mask provides the necessary volume normalization.
This is the smoothing procedure advocated by Melott \& Dominik (1993),
who tested a variety of schemes for defining smoothed density fields
from finite samples.

When we apply CONTOUR to measure the topology of isodensity surfaces
in this smoothed field, we sum the Gaussian curvature only over those
vertices whose surrounding cells all lie within the sample volume
(Weinberg 1988; Gott et al.\ 1989).  We compute the genus at 19 values 
of $\nu$, ranging from $-2.5$ to 2.5, with $\nu$ defined in terms
of the contour's enclosed fractional volume by equation~(\ref{nudef}).
In order to reduce noise in the genus curve, we set 
$G_s(\nu)$ equal to the mean of the five genus values measured at
$\nu-0.05$, $\nu-0.025$, $\nu$, $\nu+0.025$, and $\nu+0.05$;
this is similar to the boxcar local smoothing used by
Vogeley et al.\ (1994).

If we could measure the topology of the galaxy distribution from perfect
data in very large volumes, then the genus curve $g_s(\nu)$, where 
$g_s\equiv G_s/V$ is the genus per unit volume, would approach a
global average that would be independent of the details of the
data sample.  The genus curve measured in a limited volume will
differ from this global mean genus curve in part because of
random statistical fluctuations, which would average to zero in the
analysis of many independent, equivalent samples of the same size.
However, there are also effects that cause the genus curves measured
from finite galaxy redshift samples to differ systematically
from the true global mean.  The first is discreteness error, which
arises because our input galaxy distribution is a series of 
Dirac $\delta$-functions rather than a continuous field.  
Our adopted smoothing criterion, $\lambda \approx \dbar/\sqrt{2}$,
ensures enough smoothing to suppress strong discreteness distortion
(Weinberg et al.\ 1987), but the topology of the smoothed density
field may still differ from that which would be obtained by starting
from a more densely sampled galaxy distribution, with 
$\dbar \ll \sqrt{2}\lambda$.  A second class of systematic error
arises from the finite volume itself.  
Contours run into the sample boundary, so some of the holes or isolated 
pieces of the contour are not fully contained within the sample.
The CONTOUR algorithm can count ``fractional holes'' because it sums
the Gaussian curvature $K$ only over vertices contained within the sample,
but the values of $K$ along the boundary are correlated, and there can
be systematic biases whenever the summed curvature along the boundary
is a significant fraction of the summed curvature in the interior.
Another finite volume error arises because we plot $g_s$ against 
$\nu$ defined by equation~(\ref{nudef}) using the fractional volume
$f$ enclosed by the contour within the survey region.
In general this is not the same as the global value of $f$ at the
same threshold density, though it should be close if the survey volume 
is large enough to be statistically representative.
A third systematic error arises because we set the
density contrast outside the sample region to zero before smoothing
instead of to the true (but unknown) density contrast, so that
cells within one or two smoothing lengths of the boundary do not
have the correct smoothed density value.
Finally, peculiar velocities can distort the genus curve because
we compute galaxy positions from redshifts rather than true distances.
Melott et al.\ (1988) use numerical simulations to argue that peculiar
velocity effects are small, and Matsubara (1996) demonstrates this
point analytically using linear perturbation theory.

We will compare our observational results for the 1.2 Jy data
to theoretical predictions based on mock catalogs that are analyzed
in the same fashion as the data.  This approach allows a fair test 
of theoretical models regardless of the systematic effects,
but it is nonetheless valuable to know just how these influence the
observed genus curve.  We can judge this by comparing the average
mock catalog results to those obtained from the full, periodic,
densely sampled simulation cubes, which constitute effectively
perfect data.

\begin{figure}
\centerline{
\epsfxsize=6.0truein
\epsfbox[30 430 565 695]{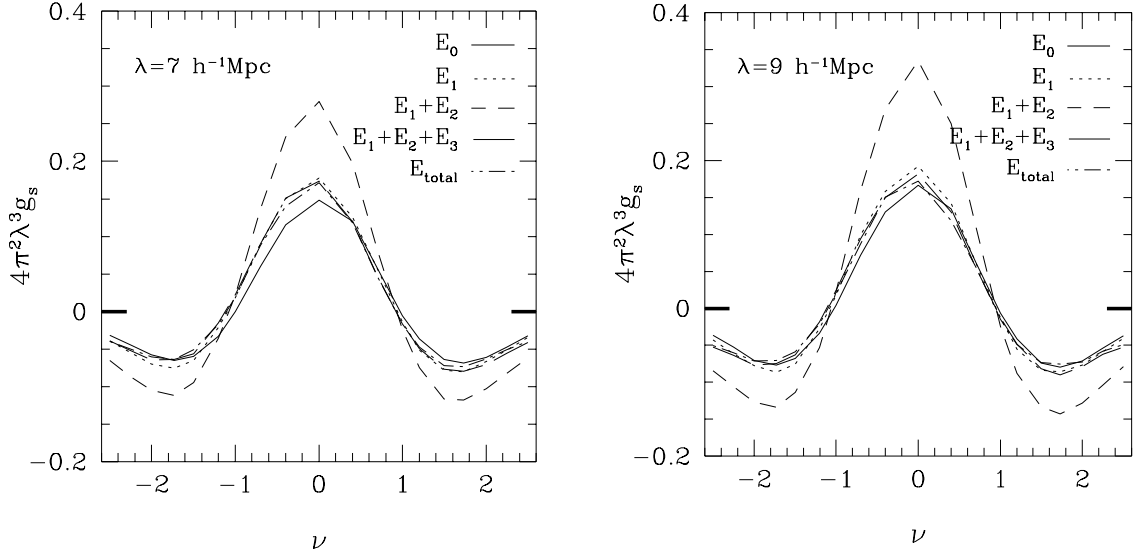}
}
\caption{\protect\small The influence of systematic biases on the
measured genus curve, for smoothing lengths $\lambda=7\hmpc$ (left)
and $\lambda=9\hmpc$ (right) and sample geometry and mean galaxy density
equivalent to those of the $\rmax=60\hmpc$ volume limited subset of the
1.2 Jy survey.  Solid lines show genus curves measured from
the effectively ``perfect'' theoretical data provided by our four CDM 
simulations, each of which is a densely sampled, $400\hmpc$, periodic cube.  
Dotted, short-dashed, long-dashed, and dot-dashed curves incorporate,
successively, discreteness effects, finite volume effects, boundary
smoothing effects, and peculiar velocity effects, as discussed in 
the text.  
\label{fig:systematic}
\normalsize}
\end{figure}

Figure~\ref{fig:systematic} illustrates this comparison for the
$\rmax=60\hmpc$ data sample using the low-$\Omega$ CDM simulations
described in \S 2, which have a power spectrum similar to that
observed for IRAS galaxies (Fisher et al.\ 1993).
The solid lines show the genus curves measured from the full,
densely sampled simulation cubes, with $\lambda=7\hmpc$ (left)
and $\lambda=9\hmpc$ (right).  We average results from the four
$400\hmpc$ cubes, and we plot $4\pi^2 \lambda^3 g_s$, where $g_s$
is the genus per unit volume, so that the expected amplitude of 
the curve is of order unity and independent of the simulation volume
itself (see eqs.~[\ref{wcurve}] and~[\ref{amplitude}]).
These curves, free of systematic errors, are labeled $E_0$.
The dotted curves are measured from the full, periodic cubes
after the galaxy distribution is randomly sampled to the mean density
of the 1.2 Jy, $\rmax=60\hmpc$ sample, 
${\overline n}_g=1.17 \times 10^{-3} h^3\;{\rm Mpc}^{-3}$.
These curves, labeled $E_1$, contain discreteness error but
no other systematic biases.  Comparing to the solid curves, we
see that discreteness raises the amplitude of the genus curve
and shifts it to the left, but with $\lambda=7$ or $9\hmpc$ the
impact is small.  Next we take the smoothed field created
from the sampled particle distribution in the periodic cubes, but
we measure the topology in volumes that have the size and geometry
of the 1.2 Jy sample, i.e., $60\hmpc$ spheres with $10^\circ$ wedges
removed.  Averaging over 512 such volumes, we obtain the dashed curves
in Figure~\ref{fig:systematic}, labeled $E_1+E_2$ because they
include both discreteness error and finite volume effects.
The latter amplify the genus curve substantially at all values of $\nu$.
However, when we smooth using only the sample volume itself,
i.e., carry out our full observational procedure but on mock catalogs
that use true distances instead of redshifts, we obtain the 
long-dashed curves $E_1+E_2+E_3$, which are much closer to the
true genus curves $E_0$.  It thus appears that the finite volume and
boundary smoothing errors largely cancel each other.
Finally, the dot-dashed
curve, labeled $E_{\rm total}$, shows the average genus curve measured
from the 512 mock catalogs in redshift space.  It is similar to
$E_1+E_2+E_3$, indicating, as expected from earlier studies,
that peculiar velocities have little effect on the genus curve.

As we shall soon see, the difference between the $E_0$ and
$E_{\rm total}$ curves in Figure~\ref{fig:systematic} is small 
compared to the random statistical fluctuations expected in 
a single volume the size of our 1.2 Jy sample.  In this sense, the
cumulative effect of the systematic errors that we have described
is small, though it is disconcerting to see that this small 
cumulative impact reflects a cancellation between two types of
errors (finite volume and boundary smoothing) that appear to be
quite substantial individually.  
After a number of tests, we remain somewhat puzzled about the nature
of the finite volume effect and the reason for its cancellation by
the boundary smoothing effect.  The systematic amplification of the
genus curve in a finite volume occurs for cubical masks and spherical
masks as well as for our IRAS masks (spheres with missing wedges),
and it occurs for Gaussian random fields as well as for N-body
models.  It gradually disappears as the survey volume becomes large
(compared to the smoothing volume), presumably because ``fractional''
holes then make a small contribution to a contour's total genus.
Note that the difference between locally and globally defined values
of $\nu$ (e.g., the fact that $\nu=0$ corresponds to the median density
within the sample instead of the true median density) can alter the
shape of the genus curve but cannot produce an overall amplification,
so it is not responsible for the effect seen here.

The systematic biases illustrated in Figure~\ref{fig:systematic} are not
significantly larger for $\lambda=7\hmpc$ than for $\lambda=9\hmpc$.
Even though the larger smoothing length suppresses discreteness more
effectively, it leads to larger finite volume effects because the
number of independent structures within a sample volume is smaller.
The random statistical fluctuations should be smaller with $\lambda=7\hmpc$
for the same reason, so we adopt this smoothing length for the
1.2 Jy, $\rmax=60\hmpc$ sample.  With a much smaller smoothing length,
discreteness effects would become excessive.  We have carried out
analysis similar to that in Figure~\ref{fig:systematic} for our
$\rmax=30\hmpc$ and $\rmax=100\hmpc$ samples, with 
$\lambda \approx \dbar/\sqrt{2}=4\hmpc$ and $12\hmpc$, respectively.
The qualitative results are similar, although the systematic biases
are somewhat stronger for these sample volumes because of the smaller
numbers of resolution elements (see Figure~\ref{fig:nres}).

\section{Random Errors and Statistical Methodology}
We ultimately wish to use the measured topology of structure in the
1.2 Jy survey to evaluate the viability of theoretical models for
the origin of this structure.  By using mock catalogs, we can ensure
that the theoretical predictions incorporate the same systematic
biases that influence the observational data.  We are then faced
with the task of deciding whether the model predictions are consistent with
the data to within the expected random errors, and whether the
data favor one theoretical model over another.  
These evaluations will have the maximum statistical power if they
are based on the likelihood $\cL\equiv P(\data | \model)$,
the conditional probability of obtaining the observed data given
the assumed model.

If the random errors in the values of $G_s$ measured at $N$ different 
values of $\nu$ were independent and Gaussian distributed, then the 
likelihood would be
\begin{equation}
\displaystyle
\cL= \exp(-{1\over 2}\chi^2_{\rm diag})\;
\prod_{i=1}^N {(2\pi\sigma_i^2)^{-1/2}},
\label{likelihood1}
\end{equation}
where
\begin{equation}
\chi^2_{\rm diag}=\sum_{i=1}^N 
{ \left[G_s^\data(\nu_i)-{\overline G}_s^\model(\nu_i)\right]^2 
\over {\sigma_i^2} },
\end{equation} 
and ${\overline G}_s^\model(\nu_i)$ and $\sigma_i^2$ are respectively
the mean and variance of the genus values obtained from the mock catalogs.
In a ``frequentist'' statistical analysis, one typically evaluates 
the acceptability of a model by asking whether its $\chi^2_{\rm diag}$ 
is ``reasonable'', i.e., whether $\chi^2_{\rm diag}/N \la 1$.
In a Bayesian analysis, the ratio of likelihoods for two models tells
one how to update the relative assessment of these models in light of the 
new data, since the ratio of models' posterior probabilities 
is equal to the ratio of their prior probabilities multiplied by
their likelihood ratio.  

\begin{figure}
\centerline{
\epsfxsize=5.0truein
\epsfbox[50 175 565 610]{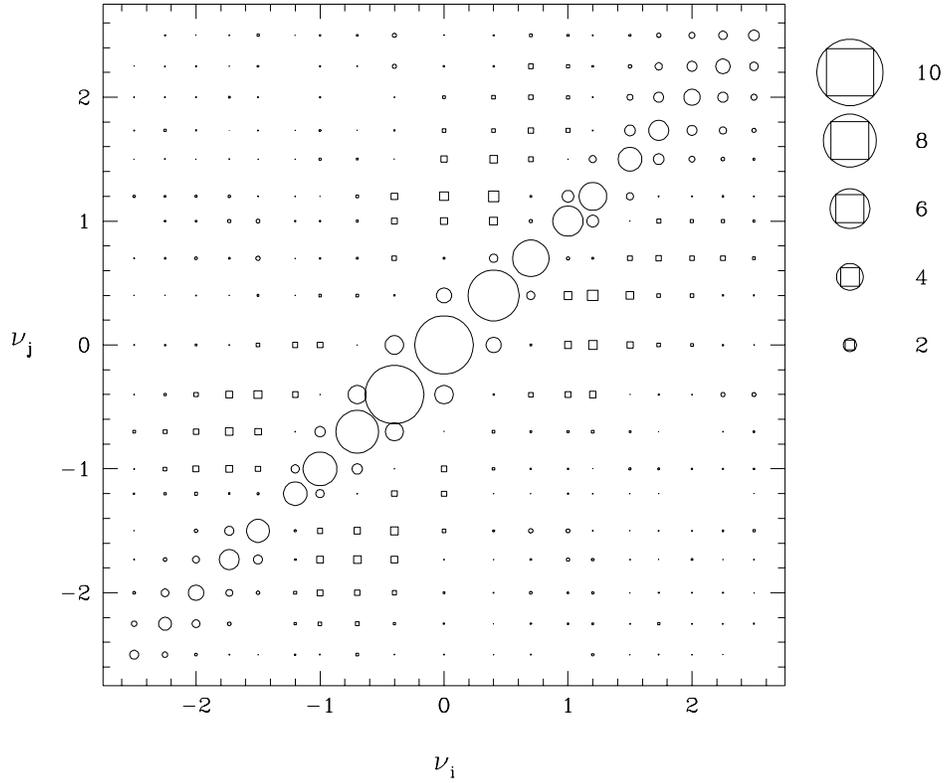}
}
\caption{\protect\small The covariance matrix $C_{ij}$ 
(eq.~[\ref{cij}]) obtained from the CDM mock catalogs with
$\rmax=60\hmpc$ and $\lambda=7\hmpc$.
The area of the symbol plotted at $\nu_i,\nu_j$ is proportional
to the magnitude of $C_{ij}$ (see scale at right), with circles and
squares representing positive and negative matrix elements, respectively.
\label{fig:covar}
\normalsize}
\end{figure}

\begin{figure}
\centerline{
\epsfxsize=5.0truein
\epsfbox[50 175 565 610]{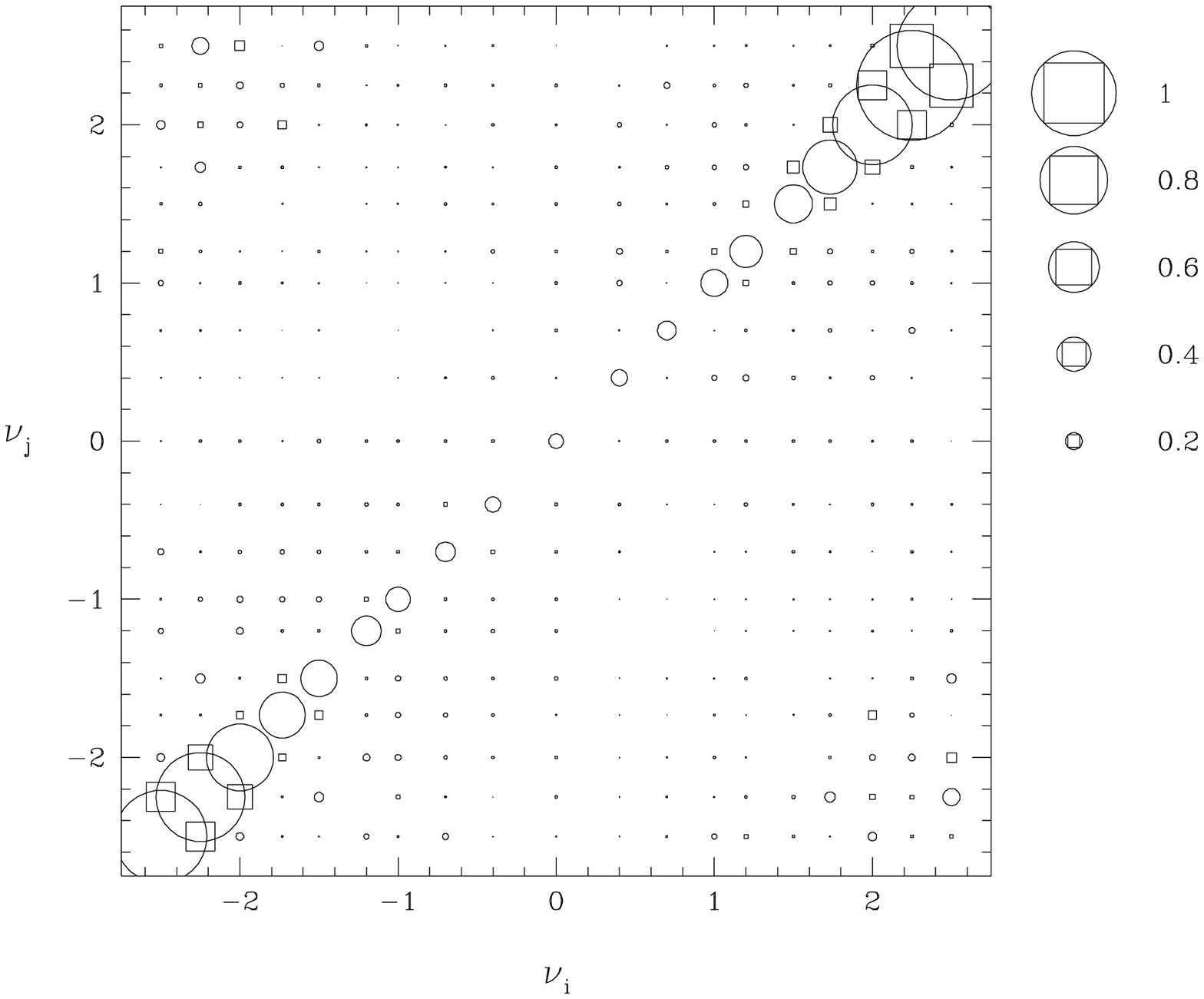}
}
\caption{\protect\small The inverse covariance matrix $C^{-1}_{ij}$
for the CDM mock catalogs, in the same format as Fig.~\ref{fig:covar}.
\label{fig:inverse}
\normalsize}
\end{figure}

The statistical evaluation of topology data is complicated by the
fact that random errors at different values of $\nu$ are {\it not}
independent.  Error correlations arise because a given volume contains
only one set of structures --- changing $\nu$ in a given volume is
not equivalent to changing $\nu$ and simultaneously moving to a different 
region of the universe to obtain an independent genus measurement.
The genus curve of an individual mock catalog is typically shifted or
amplified coherently relative to the average model prediction.
Figure~\ref{fig:covar} shows the covariance matrix,
\begin{equation}
C_{ij}=\left\langle [G_s^\model(\nu_i)-{\overline G}_s^{\model}(\nu_i)]
                    [G_s^\model(\nu_j)-{\overline G}_s^{\model}(\nu_j)]
       \right\rangle ,
\label{cij}
\end{equation}
computed from 512 mock catalogs of the $\rmax=60\hmpc$ sample drawn from
the CDM simulations, with $\lambda=7\hmpc$.
Circles and squares represent positive and negative values of $C_{ij}$,
respectively, and the area of the symbol shows the magnitude $|C_{ij}|$.  
The matrix is approximately diagonal in the sense that the largest element 
in any row $\nu_j$ is the variance $C_{jj}=\sigma_j^2$,
but there are significant correlations between the errors at neighboring
values of $\nu$, and there are anticorrelations for $|\nu_j-\nu_i| \sim 1$,
reflecting the coherent shifts of genus curves mentioned above.
The errors are largest at $\nu \approx 0$, but this is mainly because
the genus curve itself peaks here.

Even though the errors in the genus curve are not independent,
we can adopt the working hypothesis that the error distribution is a
{\it multivariate} Gaussian.  
In this case the likelihood is
\begin{equation}
\cL=(2\pi)^{-N/2}|C_{ij}|^{-1/2} \exp(-{1\over 2} \chi^2),
\label{likelihood}
\end{equation}
where
\begin{equation}
\chi^2=\sum_{i,j}^N [G_s^\data(\nu_i)-{\overline G}_s^\model(\nu_i)]
		    C^{-1}_{ij}
                    [G_s^\data(\nu_j)-{\overline G}_s^\model(\nu_j)] ,
\label{chisqr}
\end{equation}          
and $C^{-1}_{ij}$ is the inverse of the covariance matrix defined
in equation~(\ref{cij}).
If the errors were independent, then the covariance matrix would be
diagonal, $C_{ij}=\delta_{ij}\sigma_i^2$, and equation~(\ref{likelihood})
would reduce to equation~(\ref{likelihood1}).
To the extent that the error distribution is indeed multivariate Gaussian,
the quantity $\chi^2$ should follow a $\chi^2$ distribution with $N$
degrees of freedom, so we can still use the criterion $\chi^2/N \la 1$
as a frequentist evaluation of a model's success.
Even if the multivariate Gaussian assumption does not hold perfectly,
$\chi^2$ still provides a useful goodness-of-fit measure, and we can
use the mock catalogs to derive its distribution empirically.
With the Gaussian-error approximation, we can also use 
equation~(\ref{likelihood}) to compute likelihood ratios for comparisons
between models.

Figure~\ref{fig:inverse} shows the inverse covariance matrix
$C_{ij}^{-1}$ corresponding to the covariance matrix in Figure~\ref{fig:covar}.
The main impact of the correlations in $C_{ij}$ is to introduce
negative terms immediately off the diagonal in $C_{ij}^{-1}$.
From equation~(\ref{chisqr}) we see that these negative off-diagonal
terms mean that deviations of the same sign at neighboring values of
$\nu$ are ``penalized'' in the likelihood less strongly than they would be 
if we ignored the error correlations by using equation~(\ref{likelihood1}).

While our method of treating correlated errors has not been used
in previous topology analyses, it is similar to the approach used 
by Fisher et al.\ (1994b) and Cole et al.\ (1995) in their
studies of anisotropic redshift-space clustering in the 1.2 Jy survey.

\section{Results}
\subsection{Topology of the 1.2 Jy Survey}
As mentioned earlier, we perform the topology analysis at three
different smoothing scales, $\lambda=4$, 7, and $12\hmpc$.
For each scale, we use a volume limited subset of the 1.2 Jy
survey with outer radius $\rmax$ chosen so that the mean
intergalaxy separation is $\dbar \approx \lambda\sqrt{2}$.
The sample radii are $\rmax=30$, 60, and $100\hmpc$, respectively.
Since each sample is at least 4.6 times the volume of the
preceding one, and the smoothing scales themselves differ by
factors $\sim 1.7$, the genus curves obtained from these three
samples are effectively independent.

\begin{figure}
\centerline{
\epsfxsize=3.5truein
\epsfbox[30 430 290 700]{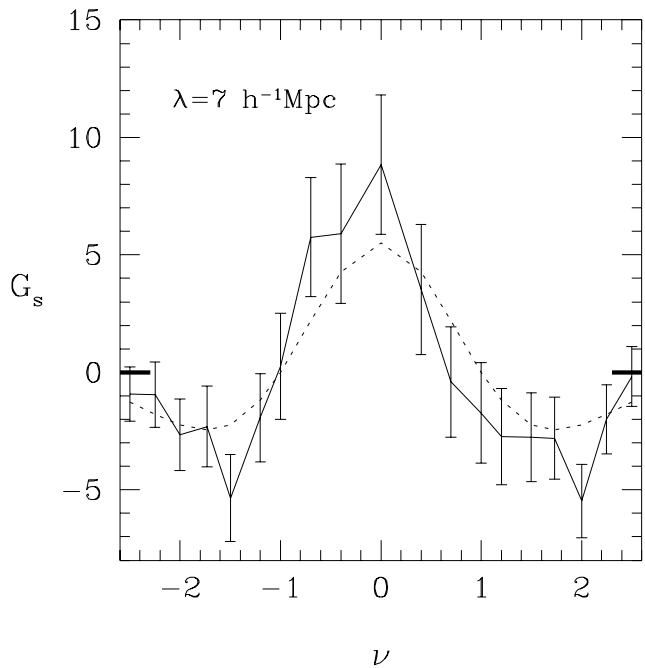}
}
\caption{\protect\small Genus curve of the $\rmax=60\hmpc$, volume
limited subset of the IRAS 1.2 Jy survey (solid line),
with a smoothing length $\lambda=7\hmpc$.
Error bars (1$\sigma$) are computed from the CDM mock catalogs.
The dotted line shows the genus curve expected for a Gaussian random
field, eq.~(\ref{wcurve}), with amplitude chosen by $\chi^2$ minimization
as described in the text.
\label{fig:iras60}
\normalsize}
\end{figure}

\begin{figure}
\centerline{
\epsfxsize=3.5truein
\epsfbox[30 430 290 700]{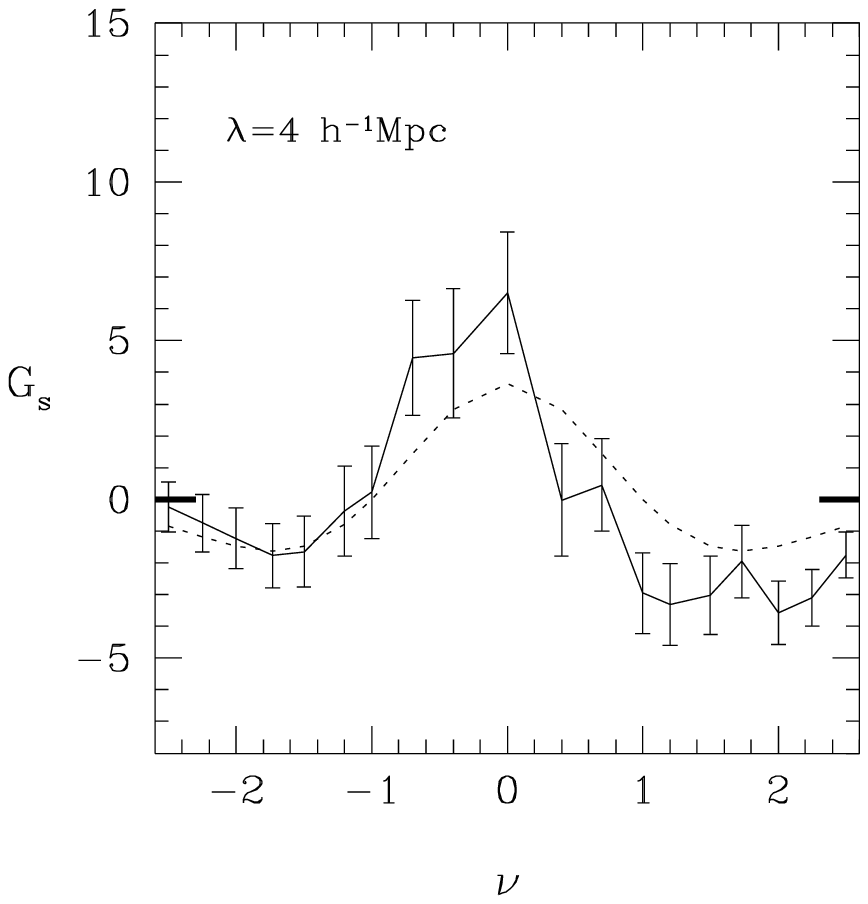}
}
\caption{\protect\small Like Fig.~\ref{fig:iras60}, but 
for the $\rmax=30\hmpc$ sample and smoothing length $\lambda=4\hmpc$.
\label{fig:iras30}
\normalsize}
\end{figure}

\begin{figure}
\centerline{
\epsfxsize=3.5truein
\epsfbox[30 430 290 700]{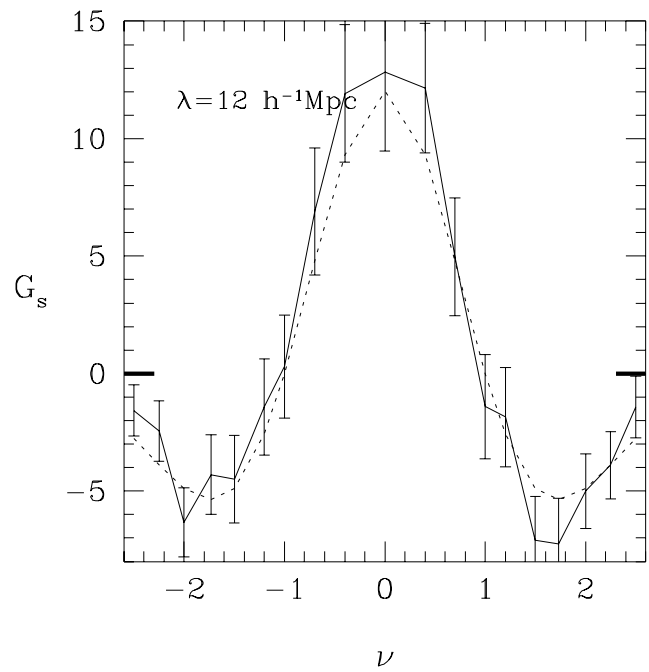}
}
\caption{\protect\small Like Fig.~\ref{fig:iras60}, but 
for the $\rmax=100\hmpc$ sample and smoothing length $\lambda=12\hmpc$.
\label{fig:iras100}
\normalsize}
\end{figure}

The $\rmax=60\hmpc$ sample has the largest number of resolution elements,
as shown in Figure~\ref{fig:nres}.
The solid line in Figure~\ref{fig:iras60} shows the genus curve of this sample.
We attach $1\sigma$ error bars computed from the CDM mock catalogs;
they are the square-roots of the diagonal elements of the covariance
matrix~(\ref{cij}) illustrated in Figure~\ref{fig:covar}.
While the size of the error bars varies from one theoretical model
to another, we have chosen a model that is known to reproduce other
clustering properties of galaxies in the 1.2 Jy survey fairly well.
Figures~\ref{fig:iras30} and~\ref{fig:iras100} show the genus
curves of the $30\hmpc$ and $100\hmpc$ samples, respectively,
with error bars computed from the CDM mock catalogs of these samples.

We will conduct a detailed comparison to theoretical models in the next
section, but as a guide in Figures~\ref{fig:iras60}--\ref{fig:iras100}
we show a genus curve with the form~(\ref{wcurve}) predicted for a Gaussian 
random field.  In each case we choose the amplitude $A$ by minimizing
$\chi^2$ using the covariance matrix computed from the CDM mock catalogs.
(A better way of choosing the amplitude would be to use the covariance
matrix for the Gaussian random field models, but as we do not have mock
catalogs for those we do not know the covariance matrix.)
The fitted amplitudes are $A=5.5$, 3.6, and 12.0 for $\rmax=60$, 30,
and $100\hmpc$, respectively, corresponding (through 
eq.~[\ref{amplitude}]) to effective power
spectral indices $n_{60}=-1.9$, $n_{30}=-1.9$, and $n_{100}=-1.0$.
At smoothing lengths $\lambda=4\hmpc$ and 
$\lambda=7\hmpc$, the observed genus curves
are shifted to the left relative to the best fitting Gaussian
field predictions.  This shift in the direction of an isolated cluster
or ``meatball'' topology has also been seen in a number of other
data samples at similar smoothing lengths (Gott et al.\ 1989; 
Moore et al.\ 1992; Park et al.\ 1992).
Yess et al.\ (1996) also find evidence for
a ``meatball'' topology in the 1.2 Jy survey from percolation analysis.
At $\lambda=12\hmpc$, the observed genus 
curve is symmetric and similar in form to the Gaussian field prediction.
Visual examination with the plotted error bars suggests that
the observed genus curve disagrees significantly with the Gaussian field curve
for $\lambda=4\hmpc$, is marginally compatible with it
for $\lambda=7\hmpc$, and matches it well for $\lambda=12\hmpc$.
This impression is borne out by the $\chi^2$ values, which are
34.0, 21.2, and 15.3 for the three smoothing lengths, respectively,
with 18 degrees of freedom (19 data points less one free fitting parameter).
These $\chi^2$ values should not be taken too literally, since they
are all computed using the CDM covariance matrix.
Still more important, the Gaussian field predictions do not include
any effects of nonlinear gravitational evolution, and they do not
include the systematic biases discussed in \S 3.

\subsection{Comparison to Models}
Our goal in this section is to test dynamically evolved models with
Gaussian initial conditions against the observed topology of the
1.2 Jy survey.  We will also consider a simple example of a model
with a non-Gaussian topology.  As discussed in \S 2, we have performed N-body
simulations starting from Gaussian initial conditions with a
$\Gamma=0.25$ CDM spectrum and power-law spectra with $n=0$, 
$n=-1$, and $n=-2$.  In all models we assume that $\Omega_0=0.3$
and that galaxies are unbiased tracers of the mass distribution.
For each of the models,
we have used these simulations to construct 512 mock 1.2 Jy catalogs
at each value of $\rmax$, and from these we 
compute the mean predicted genus curve and the covariance matrix of
the random errors.

\begin{figure}
\centerline{
\epsfxsize=6.0truein
\epsfbox[35 435 595 695]{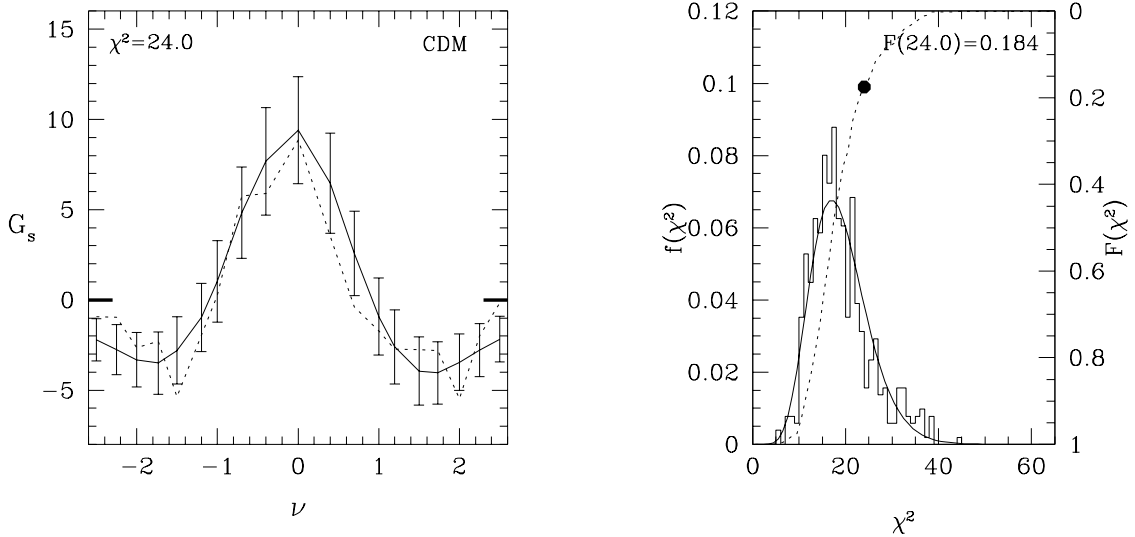}
}
\caption{\protect\small 
Comparison between the genus curve of the 1.2 Jy data
($\rmax=60\hmpc$ sample, $\lambda=7\hmpc$) and the predictions
of the low-$\Omega$ CDM model.
The left panel shows the observed genus curve (dotted line) and the
mean genus curve of the 512 mock catalogs drawn from the CDM simulation
(solid line).  Error bars show the $1\sigma$ dispersion of the
mock catalog results.  The $\chi^2$ value computed using the 
model covariance matrix is $\chi^2=24.0$.  
In the right hand panel, the solid histogram shows the distribution 
$f(\chi^2)$ of mock catalog
$\chi^2$ values relative to the mean CDM prediction.
The dotted line shows the corresponding cumulative
distribution $F(\chi^2)$.  The filled circle at $\chi^2=24.0$, 
$F(\chi^2)=0.184$ indicates that 18.4\% of mock catalogs in a CDM
universe have a $\chi^2$ value larger than that obtained for the
1.2 Jy data.  The smooth solid curve shows the distribution 
$g(\chi^2:19)$ expected for the case of a multivariate Gaussian
error distribution.
\label{fig:cdm}
\normalsize}
\end{figure}

As an example, Figure~\ref{fig:cdm} shows, in the left hand panel,
the mean genus curve of the CDM mock catalogs (solid line) for
$\lambda=7\hmpc$, $\rmax=60\hmpc$.  Error bars show the dispersion
of the mock catalog genus values at each value of $\nu$; they are
the square-roots of the diagonal elements of the covariance matrix.
The dotted line shows the observed genus curve of the $\rmax=60\hmpc$ sample,
repeated from Figure~\ref{fig:iras60}.  
Visual examination suggests that the CDM model and the IRAS data
agree fairly well given the size of the $1\sigma$ error bars,
though a ``chi-by-eye'' cannot easily take error correlations into account.

The value of $\chi^2$ from this comparison, using the CDM covariance
matrix and equation~(\ref{chisqr}), is $\chi^2=24.0$.  
We can use this value to obtain a ``frequentist'' measure of the
goodness-of-fit between the CDM prediction and the observations.
Since there are
no free parameters chosen to fit the data, the number of degrees 
of freedom is $N=19$, one for each data point.
If the CDM model were correct and the
distribution of errors were truly a multivariate Gaussian, 
the probability of getting a $\chi^2$ value this large or larger
would just be the integral of the $\chi^2$ distribution for $N$
degrees of freedom:
\begin{eqnarray}
F_G(\chi^2:N) & = & \int_{\chi^2}^{\infty} {g(t:N)dt}\nonumber \\ 
            & = & {1\over {2^{N/2} \Gamma(N/2)}} \int_{\chi^2}^{\infty}
{ t^{N/2-1}e^{-t/2}dt },
\label{fchisqr}
\end{eqnarray}
where we use the subscript $G$ on $F_G$ to denote the 
Gaussian-error assumption.
However, since we have a large number of mock catalogs available to us,
we do not have to rely on equation~(\ref{fchisqr}).
Instead, we can treat $\chi^2$ as a statistic motivated by the
Gaussian-error assumption but {\it calibrate} its distribution directly
using the mock catalogs.  The histogram in the right hand panel
of Figure~\ref{fig:cdm} shows the distribution $f(\chi^2)$, the
fraction of the CDM mock catalogs that produce this $\chi^2$ value
when compared to the mean CDM genus curve, in bins of 
width $\Delta \chi^2 = 1$.  The dotted curve shows the corresponding
cumulative distribution $F(\chi^2)$, the integral of $f(\chi^2)$.  A heavy
point is plotted at the value $F(\chi^2)=0.184$ corresponding to
the observed value of $\chi^2=24.0$.  We thus see that in the CDM
model universe, 18.4\% of random observers would find a discrepancy
with the mean CDM genus curve that is as large as or larger than
that found for the 1.2 Jy survey.  We conclude that the CDM model
does indeed yield acceptable agreement (at the $\sim 1\sigma$ level)
with the observed topology of this data sample for $\lambda=7\hmpc$.

The smooth curve in the right hand panel of Figure~\ref{fig:cdm}
shows $g(\chi^2:19)$, a $\chi^2$ distribution with 19 degrees of freedom.
This curve tracks the mock catalog histogram $f(\chi^2)$ 
remarkably well, implying that the assumption of a multivariate
Gaussian error distribution is indeed a good approximation for 
these purposes.  The value $F_G(24.0:19)=0.196$ obtained from
equation~(\ref{fchisqr}) is close to the value $F(\chi^2)=0.184$
obtained from the mock catalogs.
We have carried out similar comparisons for our other models and
other sample radii.  We find that the agreement between the mock catalog
$f(\chi^2)$ and $g(\chi^2:19)$ holds quite well in most cases,
though the mock catalog distributions tend to have somewhat longer tails
towards high $\chi^2$, so assuming Gaussian errors tends to underestimate
the probability of the most extreme events.

\begin{figure}
\centerline{
\epsfxsize=6.0truein
\epsfbox[35 165 570 695]{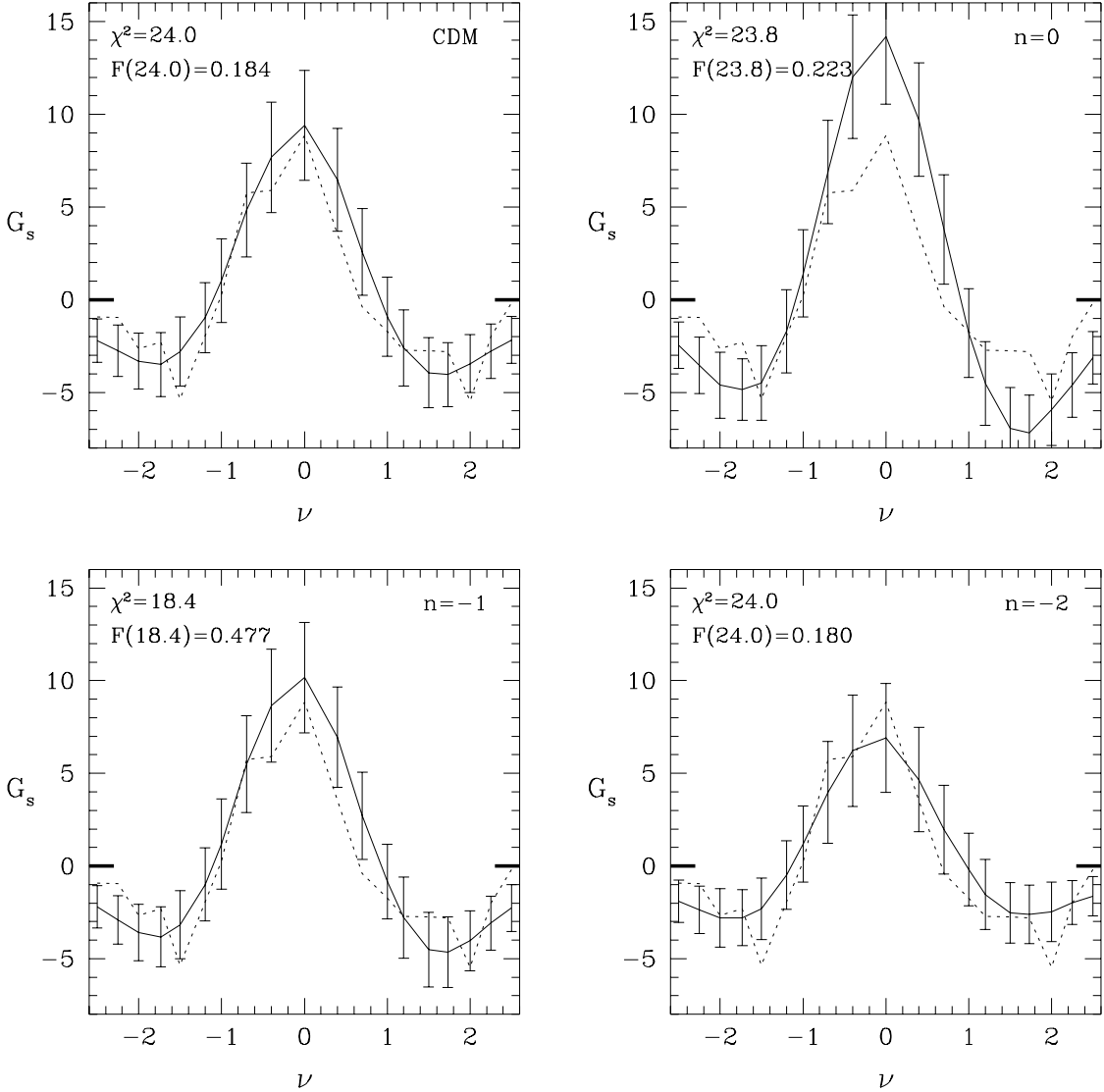}
}
\caption{\protect\small 
Comparison between the observed genus curve of the 1.2 Jy sample
for $\lambda=7\hmpc$ ($\rmax=60\hmpc$) and the 
predictions of the four N-body models, which assume Gaussian 
initial conditions with CDM, $n=0$, $n=-1$, and $n=-2$ initial
power spectra.
In each panel, the dotted line shows the observed genus curve,
and the solid line with $1\sigma$ error bars
shows the mean model prediction computed from the mock catalogs.  
The values of $\chi^2$ and $F(\chi^2)$ are listed in each panel.
\label{fig:r60}
\normalsize}
\end{figure}

\begin{figure}
\centerline{
\epsfxsize=6.0truein
\epsfbox[35 165 570 695]{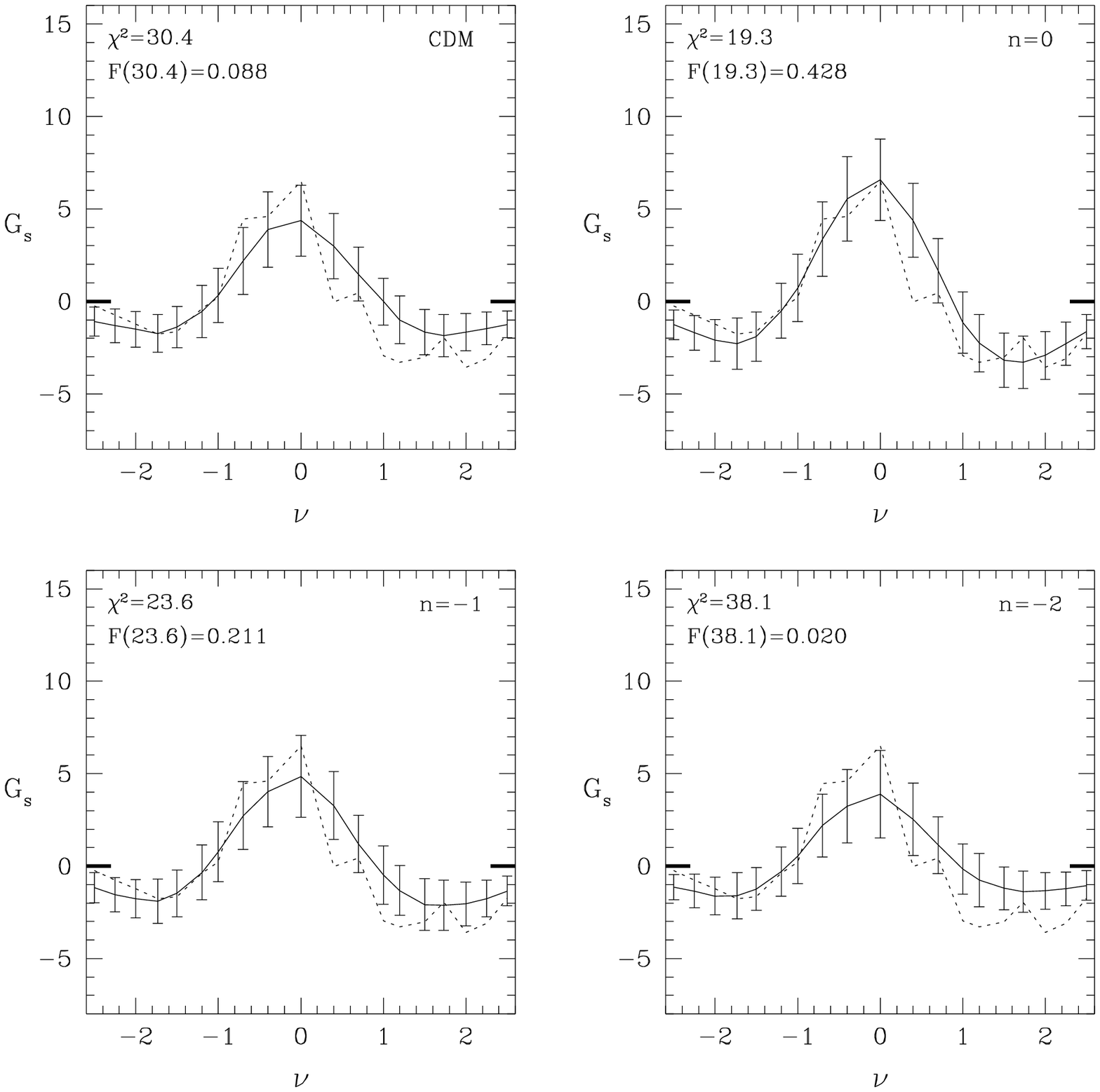}
}
\caption{\protect\small Same as Fig.~\ref{fig:r60} 
but for a smoothing scale $\lambda=4$ $h^{-1}$Mpc ($\rmax=30\hmpc$).
\label{fig:r30}
\normalsize}
\end{figure}

\begin{figure}
\centerline{
\epsfxsize=6.0truein
\epsfbox[35 165 570 695]{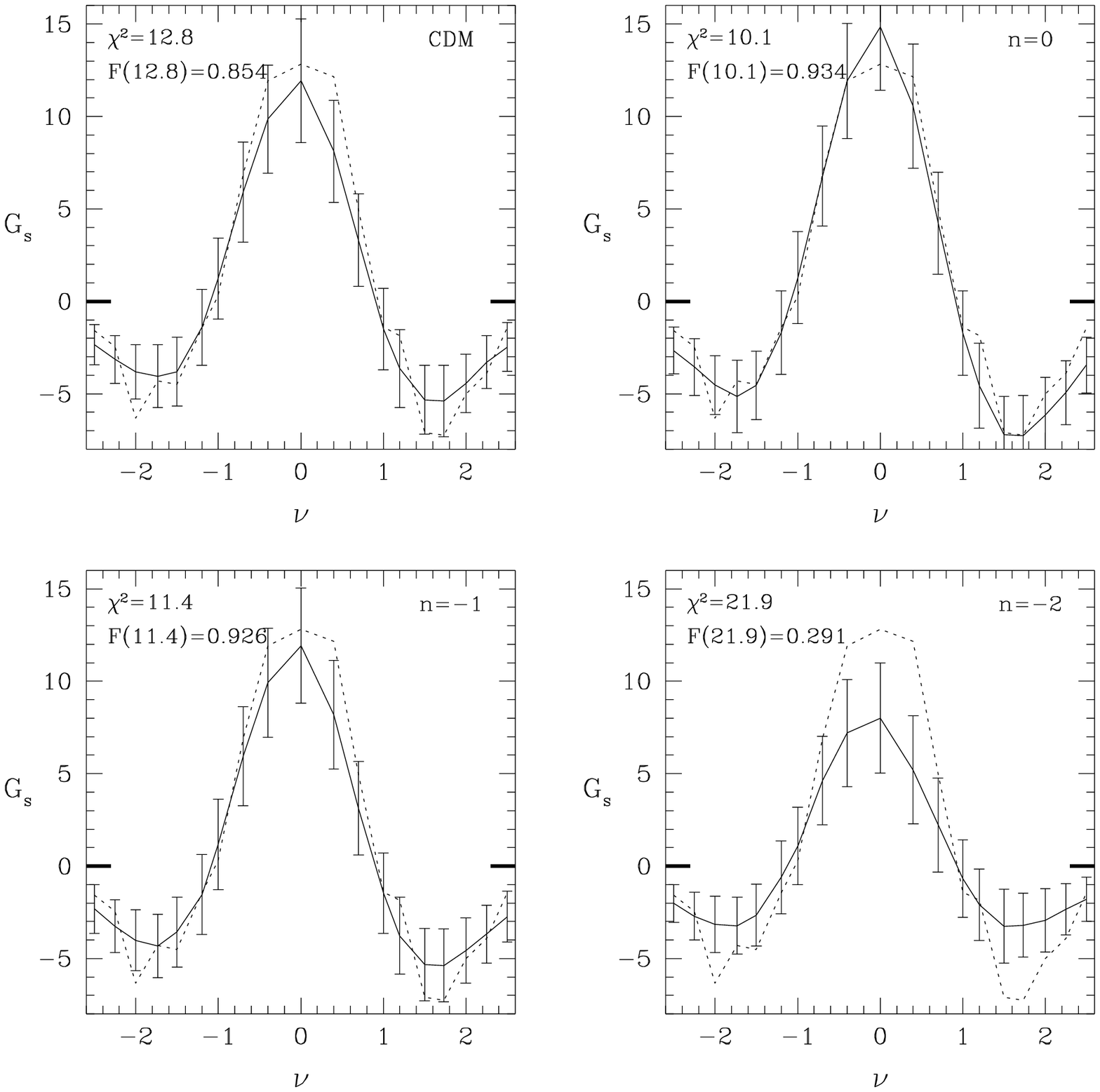}
}
\caption{\protect\small Same as Fig.~\ref{fig:r60} 
but for a smoothing scale $\lambda=12$ $h^{-1}$Mpc ($\rmax=100\hmpc$).
\label{fig:r100}
\normalsize}
\end{figure}

Figures~\ref{fig:r60}, \ref{fig:r30}, and \ref{fig:r100} show
the comparisons between our four dynamically evolved, initially
Gaussian models and the 1.2 Jy data, for 
$(\lambda,\rmax)=(7,60)$, (4,30), and (12,100), respectively. 
In each panel we show the mean predicted genus curve 
as the solid line with $1\sigma$ error bars and the observed genus
curve as the dotted line.  We also list the value of $\chi^2$
obtained using the model covariance matrix and the fraction 
$F(\chi^2)$ of mock catalogs that produce values of $\chi^2$ 
at least this large when compared to the mean model prediction.

For $\lambda=7\hmpc$ (Fig.~\ref{fig:r60}), the CDM and $n=-1$
models give similar results, both in reasonable agreement with
the observed genus curve.  The $n=0$ model predicts a higher 
amplitude genus curve, as expected given the greater amount
of small-scale power 
in its initial conditions (see eq.~[\ref{amplitude}]).
However, this model also predicts larger error bars than the CDM or
$n=-1$ models, and its $\chi^2$ of 23.8 is similar to the
CDM $\chi^2=24.0$.
A discrepancy of $\chi^2 \geq 23.8$ 
is found for $F(\chi^2)=22.3\%$ of the $n=0$ mock catalogs.
The $n=-2$ model predicts the lowest amplitude genus curve, 
and it also has $\chi^2=24.0$, with $F(\chi^2)=18.0\%$.
At $\lambda=7\hmpc$, all four models predict
an approximately symmetric genus curve, with a small asymmetry between
the topology of high and low density regions that reflects the combination
of nonlinear gravitational evolution and systematic biases in the
topology measurements.

For $\lambda=4\hmpc$ (Fig.~\ref{fig:r30}), the amplitude of the
CDM predicted genus curve is lower than that of the observed genus curve.
The comparison yields $\chi^2=30.4$, $F(\chi^2)=0.088$, indicating only
marginal compatibility between the model and the data.
The $n=0$ model predicts a higher amplitude curve that agrees fairly
well with the observations.  The $n=-1$ prediction is similar to
the CDM prediction, but it is different enough to yield better
quantitative agreement with the 1.2 Jy data.  The $n=-2$ model
predicts a genus curve whose amplitude is too low.
It is strongly contradicted by the data, with $\chi^2=38.1$
and $F(\chi^2)=0.02$.  At this smoothing length, all four
models predict a genus curve that is shifted slightly in the
direction of a ``meatball'' topology, but in all cases the
shift is smaller than that seen in the 1.2 Jy genus curve.

At $\lambda=12\hmpc$ (Fig.~\ref{fig:r100}), the observed genus
curve is quite symmetric, and it agrees well with the CDM, $n=0$,
and $n=-1$ model predictions.  Indeed, the match to the $n=0$ 
and $n=-1$ curves is
so good that in each case more than 90\% of the mock catalogs have higher
$\chi^2$ than the 1.2 Jy data.  The amplitude of the $n=-2$
genus curve is lower than observed, but for this smoothing length
the discrepancy with the data ($\chi^2=21.9$)
is not very significant, with $F(\chi^2)=29.1\%$.

\begin{figure}
\centerline{
\epsfxsize=6.0truein
\epsfbox[35 165 570 695]{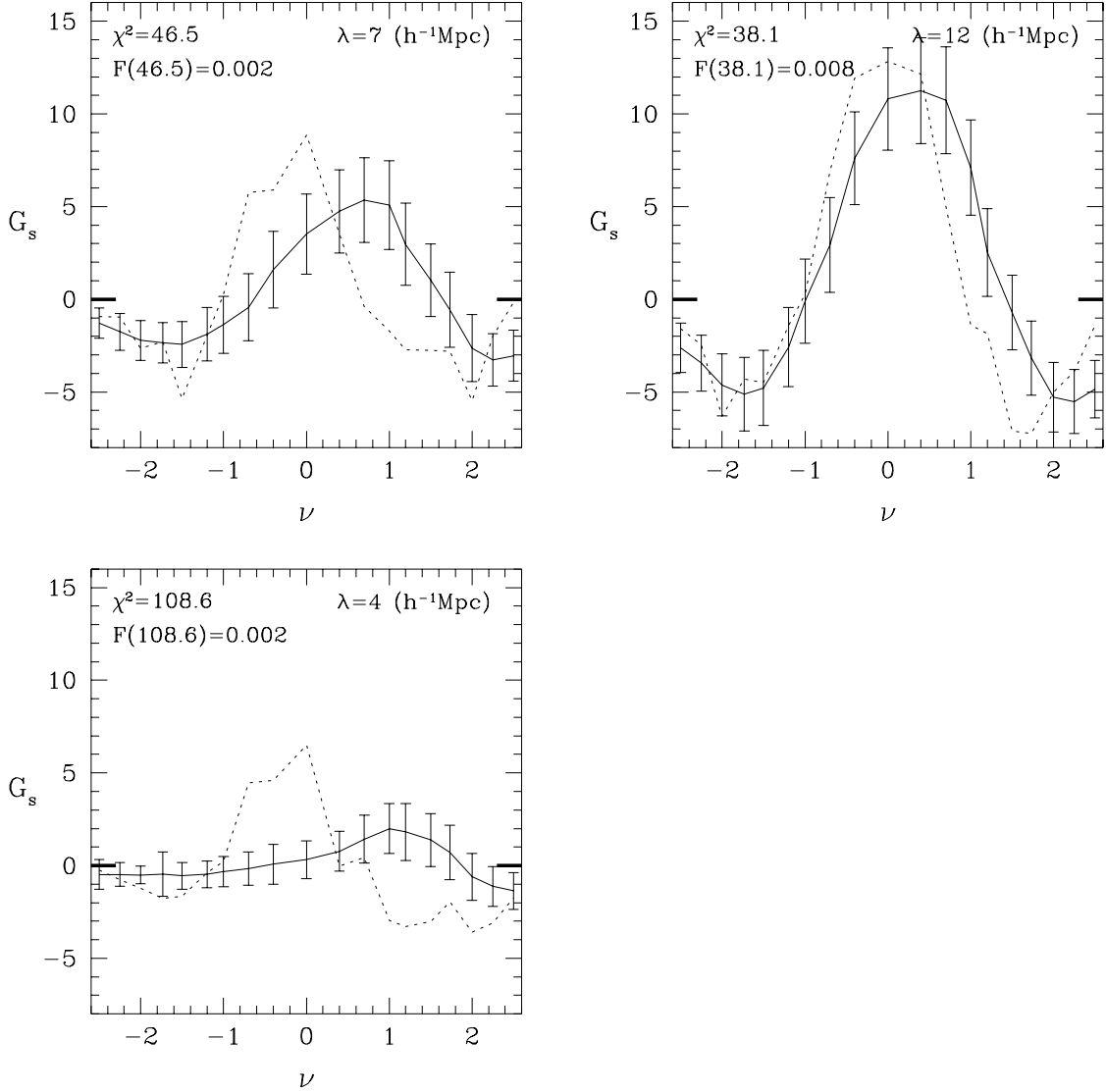}
}
\caption{\protect\small Comparison of the observed genus curves
(dotted lines) to the predictions of the Voronoi foam model
(solid lines with $1\sigma$ error bars), for the three 
smoothing lengths and corresponding sample volumes.
Values of $\chi^2$ and $F(\chi^2)$ are listed in each panel.
\label{fig:voronoi}
\normalsize}
\end{figure}

As a simple example of a model with a non-Gaussian topology, we consider
a Voronoi foam, which has often been used as a phenomenological description
of the large scale structure of the galaxy distribution
(van de Weygaert 1994 and references therein).
Starting with a Poisson distribution of particles in a box of size $300\hmpc$,
we randomly distribute ``seeds'' with a mean separation
$\dbar_s=\overline{n}_s^{-1/3}=50\hmpc$, then project each particle 
radially outward from the nearest seed until it is equidistant with
the second-nearest seed.  This procedure distributes the particles
on the walls of polygonal cells, whose faces are the perpendicular
bisector planes of neighboring seeds.
The resulting ``bubble'' topology is similar to that of structure
that evolves from initial conditions with a negatively skewed
probability distribution (Weinberg \& Cole 1992).
We generate four Voronoi foams of this sort and create mock catalogs
and compute mean predictions and covariance matrices just as we
have done for the N-body models.

Figure~\ref{fig:voronoi} shows the comparison between the Voronoi
foam model and the 1.2 Jy data for the three smoothing lengths.
On all three scales, the genus curve of the Voronoi foam shows the
rightward shift characteristic of a bubbly (as opposed to spongelike)
distribution.  In every case, the $\chi^2$ value is very large,
greater than that in 99\% or more of the mock catalogs.

Table~1 summarizes the results of these comparisons, listing the
values of $\chi^2$ and $F(\chi^2)$ for each combination of model and
smoothing length.  We can directly compare the ability of these
models to account for the topology of the 1.2 Jy data using a likelihood
ratio test.  For this purpose, we will assume that the error distribution
is indeed multivariate Gaussian, so that the relative likelihood
of obtaining the observed genus curve in models~A and~B is,
from equation~(\ref{likelihood}),
\begin{equation}
\frac{\cL_A}{\cL_B} =
    \frac{|C_{ij,B}|^{1/2}}{|C_{ij,A}|^{1/2}} 
    \exp\left[-\frac{1}{2}(\chi_A^2-\chi_B^2)\right].
\label{lratio}
\end{equation}
Because the genus curves for the three smoothing lengths are
effectively independent (see \S 5.1), the overall likelihood ratio
of two models is simply the product of their likelihood ratios
for the three different samples.
Note that the covariance matrix determinants enter into the likelihood
ratio as well as the $\chi^2$ values themselves: for instance, if two
models have the same $\chi^2$ value but one predicts smaller random
errors, then the model that makes the tighter prediction is preferred.
The determinant ratio factor can be quite significant, as one can
easily see for the case of uncorrelated errors 
($C_{ij}=\delta_{ij}\sigma_{i}^2)$.  If one model predicts error
bars that are consistently 20\% larger at each of the $N=19$ $\nu$ values, 
then the determinant ratio factor is $1.2^N\approx 32$ at each 
smoothing length.

The likelihood ratios (relative to the CDM model) are listed in the
last column of Table~1.  We have already seen from the $\chi^2$
test that the CDM, $n=0$, and $n=-1$ models are separately
compatible with the observed topology of the three volume limited
1.2 Jy samples at the $>5\%$ level.
The likelihood ratios, which combine the information from all three
samples, show that the $n=-1$ model (advocated on other grounds as
early as Gott \& Rees 1975) is the most successful overall, with 
a likelihood $\sim 90$ times higher than that of the CDM model.
While the $n=0$ model produces lower $\chi^2$ values than the CDM
model at each smoothing length, its overall likelihood is only
about half that of the CDM model because of its systematically larger
error bars.  The $n=-2$ model is strongly disfavored by the data,
with a likelihood ratio of $0.002$ relative to CDM and $2.2 \times 10^{-5}$
relative to the $n=-1$ model.
We have also investigated power-law models with $\Omega_0=1$ instead
of $\Omega_0=0.3$ (but otherwise identical to our standard power-law
models), and although the predicted genus curves are not radically 
different in the high density models, the statistical discrepancies
with the observations are substantially larger for $n=-1$ and $n=-2$
and slightly smaller for $n=0$.
The Voronoi foam has a formal likelihood ratio of $4.4 \times 10^{-26}$,
and while the assumption of Gaussian errors surely breaks down
at this level, it is clear that the observed topology of the 1.2 Jy
survey is inconsistent with the ``bubble'' topology of this model.

\begin{table}[h]
\tablecaption{Model $\chi^2$ values and likelihoods}
\centerline{
\begin{tabular}{ c c c c c } \tableline\tableline
& & $\lambda\; (\rmax)$ & & \\ \cline{2-4}
Model  &7 (60)   &4 (30)     &12 (100)   &$\cL/\cL_{\rm CDM}$ \\
\tableline
CDM         &24.0 (0.184)  &30.4 (0.088)     &12.8 (0.854)   &1.00     \\
$n=0$       &23.8 (0.223)  &19.3 (0.428)     &10.1 (0.934)   &0.57     \\
$n=-1$      &18.4 (0.477)  &23.6 (0.211)     &11.4 (0.926)   &91.8     \\
$n=-2$      &24.0 (0.180)  &38.1 (0.020)     &21.9 (0.291)   &0.002\\
Voronoi     &46.5 (0.002)  &108.6(0.002)     &38.1 (0.008)   &4.4 $\times$ 
							      10$^{-26}$\\
\tableline
\end{tabular}}
\caption{\protect\small
Summary of comparisons between models and the observed genus curves.
Columns 2, 3, and 4 list the values of $\chi^2$ and $F(\chi^2)$ for
smoothing lengths $\lambda=7$, 4, and $12\hmpc$, respectively.
Column 5 lists the overall likelihood of the model relative to the
likelihood of the CDM model. }
\end{table}

\section{Summary}
We have measured the topology of the galaxy distribution in the
IRAS 1.2 Jy redshift survey using the methods of GMD, GWM, and
Gott et al. (1989).  We consider three volume limited
subsets of the data, with limiting radii $\rmax=30$, 60, and $100\hmpc$,
analyzed with corresponding smoothing lengths $\lambda=4$, 7, and $12\hmpc$.
We use mock catalogs drawn from cosmological
N-body simulations in order to derive theoretically predicted genus curves
and to study the systematic and random errors expected in samples of
this size.  Our principal conclusions are:
\\
(1) In tests on mock catalogs from low-$\Omega$ CDM simulations,
the net systematic error in volumes the size of our 1.2 Jy subsamples
is small compared to the random errors.  However, this small net
error involves a cancellation between the systematic effects of
measuring the genus in a volume that contains few independent structures
and the effects of smoothing only with that portion of the smoothing
window that lies within the sample volume.
\\
(2) The covariance matrix of random errors in the genus curve is
predominantly diagonal, but there are significant correlations
in the errors that should be taken into account when assessing 
theoretical models.  To a reasonable approximation, $\chi^2$
(including covariances) is distributed as it would be if the errors
followed a multivariate Gaussian distribution, though extreme values
of $\chi^2$ are more common than they would be for purely Gaussian statistics.
\\
(3) With $\lambda=12\hmpc$, the genus curve of the 1.2 Jy data has
a symmetric form similar to that predicted for a Gaussian random field.
For $\lambda=7\hmpc$ and $4\hmpc$, the observed genus curves are 
increasingly shifted in the direction of a ``meatball'' topology.
\\
(4) Taken individually, the three observed genus curves are 
consistent at the $>5\%$ level with the topology of
dynamically evolved N-body models that have Gaussian initial conditions
with low-$\Omega$ CDM ($\Gamma=0.25$), $n=0$, or $n=-1$ power spectra.
Combining all three data sets, the $n=-1$ model is the most successful
overall, with a likelihood ratio of 91.8 relative to CDM and 161 relative
to $n=0$.
\\
(5) The observed genus curves are inconsistent with an $n=-2$,
initially Gaussian model, which produces structure that is excessively
coherent and, consequently, genus curves whose amplitudes are too low.
The observed genus curves are strongly inconsistent with a Voronoi
foam model, which, on account of its ``bubble'' topology, predicts
genus curves that are systematically shifted towards higher densities,
in the opposite direction from the observed shifts.

Our conclusions about the shapes of observed genus curves and
their consistency or inconsistency with various theoretical models
are similar to those drawn from a number of other topological
studies of the galaxy distribution (Gott et al.\ 1989; 
Moore et al.\ 1992; Park et al.\ 1992; Colley 1996).
They are somewhat at odds with the results of Vogeley et al. (1994),
who found that genus curves measured from the extended CfA redshift
survey showed shifts in the direction of a bubble topology and
were inconsistent with the genus curves predicted by CDM N-body models.
The differences could reflect systematic differences in the structure
traced by optical and IRAS galaxies, differences in the details of
the topology analysis, or the somewhat greater statistical power of the CfA 
data set at the short smoothing lengths ($\lambda \sim 5\hmpc$) where the
differences are most pronounced.

Earlier topology studies, recognizing the problem of correlated errors
in the genus curve, have developed ``meta-statistics'' that characterize
the overall shape of the genus curve (e.g., amplitude, asymmetry, width)
and used these to assess the compatibility of models with the
observations (see, e.g., Vogeley et al.\ 1994).
Here we have taken the more direct approach of measuring the error
covariance matrix from mock catalogs and incorporating it into
model assessments.  We compute $\chi^2$ values that include the
error covariance (eq.~[\ref{chisqr}]) and calculate the distribution
of $\chi^2$ from mock catalogs, in order to get an absolute,
frequentist assessment of a model's goodness-of-fit.
The distribution of $\chi^2$ values in the mock catalogs implies
that the multivariate Gaussian approximation describes the error
distribution quite well, failing only in the extreme tails.
We perform likelihood ratio comparisons between theoretical models
to assess their relative ability to account for the observed topology
data, making use of the Gaussian-error approximation.
The advantages of our approach are that it has a clear statistical
motivation, it provides a natural path for combining information from
independent data samples, 
and, to the extent that the Gaussian-error approximation holds,
it makes the best possible use of the data because it is based
directly on the likelihood.  The disadvantage is that a high $\chi^2$
or low likelihood value says nothing in itself about how the model
prediction and the data disagree.  Thus, the likelihood approach used
here is the most statistically powerful way to assess and compare
models, but measures like those of Park et al.\ (1992) and
Vogeley et al.\ (1994) may be useful for quantifying the nature
of discrepancies between theory and observation.
Our approach to handling correlated errors is similar to that used
by Fisher et al.\ (1994) and Cole et al.\ (1995) in studies of
redshift space distortions of the correlation function and power spectrum,
and it can be adapted to many other problems in which error correlations
are important but computable.  It should be especially useful for
topological analyses of future large galaxy redshift surveys,
such as the Anglo-Australian 2dF survey and the Sloan Digital Sky Survey,
which will yield much more stringent tests of the
hypothesis that structure in the universe formed from Gaussian
primordial fluctuations.

\acknowledgments
We are grateful to Michael Strauss and Karl Fisher for providing us
with the 1.2 Jy redshift data in electronic form and for numerous
helpful discussions on large scale structure.
We thank Richard Gott, Adrian Melott, and Andrew Hamilton for
helpful discussions on topological analysis of redshift surveys
over the course of many years.
We acknowledge support from NASA Grants
NAG5-3111 and NAG5-2759 and NSF Grant AST95-29120.


\begin{references}

\reference {} Adler, R. J., 1981, The Geometry of Random Fields, Wiley, 
New York
\reference {} Bardeen, J.M., Bond, J.R., Kaiser, N., \& Szalay, A.S. 1986, 
MNRAS, 304, 15
\reference {} Bernardeau, F. 1992, \apj, 392, 1
\reference {} Bernardeau, F., \& Kofman, L. 1995, \apj, 443, 479
\reference {} Bouchet, F.R., Strauss, M.A., Davis, M., Fisher, K.B., 
Yahil, A., \& Huchra, J.P. 1993, \apj, 417, 36
\reference {} Cole, S., Fisher, K.B., Weinberg, D.H., 1995, \mnras, 275, 515
\reference {} Coles, P., Davies, A. G., \& Pearson, R. C. 1996, MNRAS, in 
press, astro-ph/9603139
\reference {} Coles, P., \& Plionis, M. 1991, \mnras, 250, 75
\reference {} Colley, W. N. 1996, \apj, submitted, astro-ph/9612106
\reference {} Colley, W.N., Gott, J.R., \& Park, C. 1996, \mnras, submitted,
astro-ph/9601084
\reference {} Davis, M., Nusser, A., \& Willick, J. 1996, \apj, 473, 22S
\reference {} Doroshkevich, A.G. 1970, Astrophysica, 6, 320
\reference {} Efstathiou, G., Bond, J.R., \& White, S.D.M. 1992, MNRAS, 258, 1
\reference {} Fisher, K.B., Davis, M., Strauss, M.A., Yahil, A., 
\& Huchra, J.P. 1993, \apj, 402, 42
\reference {} Fisher, K.B., Davis, M., Strauss, M.A., Yahil, A., \& 
Huchra, J.P. 1994a, \mnras, 266, 50
\reference {} Fisher, K.B., Davis, M., Strauss, M.A., Yahil, A., \& 
Huchra, J.P. 1994b, \mnras, 267, 927
\reference {} Fisher, K.B., Huchra, J.P., Strauss, M.A., Davis, M.,
Yahil, A., \& Schlegel, D. 1995, \apjs, 100, 69
\reference {} Fry, J.N. 1984, \apj, 279, 499
\reference {} Gott, J.R., Mao, S., Park, C., \& Lahav, O. 1992, \apj, 385, 26
\reference {} Gott, J.R., Melott, A.L., \& Dickinson, M., 1986, \apj, 306, 341 
(GMD)
\reference {} Gott, J.R., Miller, J., Thuan, T.X., Schneider, S.E., Weinberg,
D.H., Gammie, C., Polk, K., Vogeley, M., Jeffrey, S., Bhavsar, S.P., Melott, 
A.L., Giovanelli, R., Haynes, M., Tully, B.R., \& Hamilton, A.J.S, 1989, \apj, 
340, 625 
\reference {} Gott, J.R., \& Rees, M. J. 1975, A\&A, 45, 365
\reference {} Gott, J.R., Weinberg, D.H., \& Melott, A.L., 1987, \apj, 319, 1 
(GWM)
\reference {} Hamilton, A.J.S., Gott, J.R., \& Weinberg, D.H., 1986, \apj, 
309, 1
\reference {} Juszkiewicz, R., Bouchet, F.R., \& Colombi, S. 1993, \apj, 
412, L9
\reference {} Juszkiewicz, R., Weinberg, D.H., Amsterdamski, P., Chodorowski, 
M., \& Bouchet, F.R. 1995, \apj, 442, 39
\reference {} Kogut, A., Banday, A.J., Bennett, C.L., Gorski, K.M.,
Hinshaw, G., Smoot, G.F., \& Wright, E.L. 1996, \apj, 464, L29
\reference {} Matsubara, T. 1994, ApJ, 434, L43
\reference {} Matsubara, T. 1996, \apj, 457, 13
\reference {} Matsubara, T. \& Suto, Y. 1996, ApJ, 460, 51
\reference {} Melott, A. L. 1990, Phys Rep, 193, 1
\reference {} Melott, A.L., \& Dominik, K.G. 1993, ApJS, 86, 1, 1
\reference {} Melott, A.L., Weinberg, D.H., \& Gott, J.R. 1988, \apj, 328, 50
\reference {} Moore, B., Frenk, C.S., Weinberg, D.H., Saunders, W., Lawrence, 
A., Ellis, R.S., Kaiser, N., Efstathiou, G., \& Rowan-Robinson, M. 1992,
\mnras, 256, 477
\reference {} Park, C. 1990, PhD thesis, Princeton University
\reference {} Park, C. \& Gott, J.R. 1991, ApJ, 378, 457
\reference {} Park, C., Gott, J.R., \& da Costa, L.N. 1992, \apj, 392, L51
\reference {} Park, C., Gott, J. R., Melott, A.L., \& Karachentsev, I.D. 1992, 
\apj, 387, 1
\reference {} Plionis, M., Valdardini, R., \& Coles, P. 1992, \mnras, 258, 114
\reference {} Protogeros, Z.A.M., Scherrer, R.J., 1997, \mnras, in press, 
astro-ph/9603155
\reference {} Rhoads, J.E., Gott, J.R., \& Postman, M. 1994, \apj, 421, 1
\reference {} Shandarin, S. F. \& Zel'dovich, Ya. B. 1983, Comm Astrophys, 10, 33
\reference {} Smoot, G.F., Tenorio, L., Banday, A.J., Kogut, A., Wright, E.L.,
Hinshaw, G., \& Bennett, C.L. 1994, \apj, 437, 1
\reference {} Strauss, M. A., Huchra, J. P., Davis, M., Yahil, A.,
Fisher, K. B., \& Tonry, J. 1992, \apjs, 83, 29
\reference {} Van de Weygaert, R. 1994, A\&A, 213, 1
\reference {} Vogeley, M.S., Park, C., Geller, M.J., Huchra, J.P., \& Gott, 
J.R.  1994, \apj, 420, 525
\reference {} Weinberg, D. H. 1988, PASP, 100, 1373
\reference {} Weinberg, D. H., \& Cole, S. 1992, MNRAS, 259, 652
\reference {} Weinberg, D.H., Gott, J.R., \& Melott, A.L. 1987, \apj, 321, 2
\reference {} Yess, C., Shandarin, S., \& Fisher, K. 1997, \apj, 474, 553

\end{references}
\end{document}